\documentstyle[aps,epsf]{revtex}

\newcommand{\beq}{\begin{equation}}\newcommand{\beqa}{\begin{eqnarray}}
\newcommand{\eeq}{\end{equation}}\newcommand{\eeqa}{\end{eqnarray}}

\newcommand{\aout}{a_{\rm o}}
\newcommand{\rmin}{{\rm (in)}}\newcommand{\rmout}{{\rm (out)}}
\begin{document}
\twocolumn[
\begin{center}
{\large\bf Particle creation near the chronology horizon$^\dagger$}

\vskip12pt
Sergey V. Sushkov

{\it Department of Geometry, Kazan State Pedagogical University,

Mezhlauk 1 st., Kazan 420021, Russia

e-mail: sushkov@kspu.ksu.ras.ru}

\vskip6pt
Submitted to Physical Review D, January, 1998

\vskip12pt
\parbox{14cm}{\small
We investigate the phenomenon of particle creation of the massless
scalar field in the model of spacetime in which, depending on the
model's parameter $\aout$, the chronology horizon could be formed. The
model represents a two-dimensional curved spacetime with the topology
$R^1 \times S^1$ which is asymptotically flat in the past and in the
future. The spacetime is globally hyperbolic and has no causal
pathologies if $\aout<1$, and closed timelike curves appear in the
spacetime if $\aout\ge 1$. We obtain the spectrum of created particles
in the case $\aout<1$.  In the limit $\aout \to 1$ this spectrum gives
the number of particles created into mode $n$ near the chronology
horizon.  The main result we have obtained is that the number of scalar
particles created into each mode as well as the full number of
particles remain finite at the moment of forming of the chronology
horizon.

} 

\end{center}
\vskip12pt
]
\section{Introduction}
In 1988 Morris, Thorne, and Yurtsever \cite{MTY} had demonstrated how
one could manufacture closed timelike curves in a spacetime containing
two relatively moving traversable wormholes. That work kindled
considerable interest in the question whether it is possible, in
principle, to construct a ``time machine'' --- i.e. whether, by
performing operations in a bounded region of an initially ``ordinary''
spacetime, it is possible to bring about a ``future'' in which there
will be closed timelike curves. So far, in spite of many-year attempts
to answer this question, there does not exist full and clear
understanding of the problem. On the first stage of investigations, it
seemed that the quantum field theory could give a mechanism which would
be able to protect chronology. Hawking has suggested the chronology
protection conjecture which states that the laws of physics will always
prevent the formation of closed timelike curves
\cite{Hawking1,Hawking2}. His arguments were based on the
supposition that the renormalized vacuum expectation values of a
stress-energy tensor of a quantum field must diverge at the
chronology horizon which separates a region with closed timelike
curves from a region without them.
Various attempts at proving Hawking's conjecture have been made
\cite{Refs}, culminating in singularity theorems of Kay,
Radzikowski, and Wald \cite{KRW}.

However, in recent time, a number of examples of configurations with
the bounded renormalized stress-energy tensor near the chronology
horizon has been given \cite{Krasnikov,Sushkov1,Sushkov2,Visser1,Li}.
For instance, in our works \cite{Sushkov1,Sushkov2} it was shown
that, in the case of automorphic fields, there exists a specific
choice of quantum state for which the renormalized stress-energy
tensor vanishes at the chronology horizon (in fact it is zero in the
whole spacetime); Krasnikov \cite{Krasnikov}
exhibited several 2D models, and Visser \cite{Visser1} presented a 4D
model of spacetime -- the ``Roman ring'' of traversable wormholes --
for which the vacuum polarization can be made arbitrarily small all
the way to the chronology horizon; in addition, Li \cite{Li} showed
that the renormalized stress-energy tensor may be smoothed out by
introducing absorption material, such that the spacetime with a time
machine may be stable against vacuum fluctuations.

These examples indicate that it is impossible to prove the chronology
protection conjecture taking into account only effects of the vacuum
polarization. Recently Visser \cite{Visser2,Visser1} has given
arguments that the solving of the problem of chronology protection is
impossible within the context of semi-classical theory of gravity and
requires a fully developed theory of quantum gravity.

Nevertheless in this work we remain in the framework of semi-classical
quantum gravity. We shall discuss such the aspect of the quantum field
theory in spacetimes with the time machine which was so far not
investigated. Up to this time, various aspects of vacuum polarization
near the chronology horizon have mainly been investigating. Here we
consider dynamics of the chronology horizon forming.  As is known,
time-dependent processes in the presence of quantum fields are
accompanied by a pair creation. So one may expect that particles of 
quantized fields will be created in the process of the chronology
horizon forming. The aim of this work is to determine the spectrum of
particles and try to answer the question: Can the particle creation
stop the chronology horizon forming?

We use the units $c=G=\hbar=1$ throughout the paper.
\section{A model of spacetime} \label{toy}
Here we shall discuss a model of spacetime in which the chronology
horizon is being formed. Let us consider a strip
$\{\eta\in(-\infty,+\infty)$, $\xi\in[0, L]\}$ on the $\eta$-$\xi$
plane and assume that the points on the bounds $\gamma^-$:~
$\xi=0$ and $\gamma^+$:~$\xi=L$ are to be identified:
$(\eta,0)\equiv(\eta,L)$. After this procedure we obtain a manifold
$\overline M$ with a topology of cylinder: $R^1\times S^1$. Introduce on
this manifold the metric
\beq\label{M}
ds^2=d\eta^2+2a(\eta)d\eta d\xi-(1-a^2(\eta))d\xi^2,
\eeq
where $a(\eta)$ is a monotonically increasing function of $\eta$ which
has the following asymptotic behavior:
\beq\label{a_asymp}
\begin{tabular}{lcl}
$a(\eta)\to 0$ & \quad \rm if \quad & $\eta\to-\infty,$  \\
$a(\eta)\to a_{\rm o}$ & \quad \rm if \quad & $\eta\to+\infty,$
\end{tabular}
\eeq
where $a_{\rm o}$ is some constant. Further we shall call the manifold 
$\overline M$ as the spacetime $\overline M$ with the metric (\ref{M}).
Note that the metric's coefficients in Eq.(\ref{M}) do not depend on
$\xi$, so they
themself and their derivatives are taking the same values at the points
$(\eta,0)$ and $(\eta,L)$. Hence, the internal metrics and external
curvatures at both lines $\gamma^-$ and $\gamma^+$ are identical. This
guarantees the regularity of the spacetime $\overline M$.

The metric (\ref{M}) describes a curved spacetime which is
asymptotically flat in the past, $\eta\to-\infty$ (the ``in-region''),
and in the future, $\eta\to\infty$ (the ``out-region'').  Really, as
follows from (\ref{a_asymp}), if $\eta\to-\infty$, the metric (\ref{M})
takes exactly the Minkowski form
\beq\label{past}
ds^2=d\eta^2 - d\xi^2.
\eeq
In the future, $\eta\to\infty$, the metric (\ref{M}) reads
\beq\label{future}
ds^2=d\eta^2+2a_{\rm o}d\eta d\xi-(1-a_{\rm o}^2)d\xi^2,
\eeq
and takes the Minkowski form
\beq
ds^2=dt^2 - dx^2
\eeq
in new `Minkowski' coordinates
\beq
t=\eta+a_{\rm o}\xi,\quad x=\xi.
\eeq

Note that the spacetime $\overline M$ could be considered as the factor
space: $\overline M=M/{\cal R}$. Here $M$ is an universal covering spacetime
for $\overline M$, in our case it is the whole plane $(\eta,\xi)$,
and $\cal R$ is the equivalence relation:
\beq\label{ident}
(\eta,\xi+L) \equiv (\eta,\xi)
\eeq

Consider now the causal structure of the spacetime $\overline M$. With this
aim we have to define null (lightlike) curves which form a light cone at
a point. The equations for null curves could be found from the condition
$ds^2=0$. There exist exactly two null curves which go via each point
of the spacetime and their equations read
\beq\label{cone}
\begin{tabular}{l}
$\displaystyle \xi+\int^\eta\frac{d\tilde\eta}{1+a(\tilde\eta)}=const,$
\\[12pt]
$\displaystyle \xi-\int^\eta\frac{d\tilde\eta}{1-a(\tilde\eta)}=const.$
\end{tabular}
\eeq
As follows from the asymptotical properties (\ref{a_asymp}) of
$a(\eta)$ in the in-region the equations (\ref{cone}) take the
form:
\beq\label{conein}
\eta+\xi=const,\quad \eta-\xi=const,
\eeq
and in the out-region they are
\beq\label{coneout}
\eta+(1+a_{\rm o})\xi=const,\quad
\eta-(1-a_{\rm o})\xi=const.
\eeq
Analysing of the equations (\ref{cone}) together with their
asymptotical forms (\ref{conein}) and (\ref{coneout}) reveals the
following qualitative picture: The light future cone
in the past (see Eqs.(\ref{conein})) has the angle $\alpha_1=90^\circ$
and has no inclination (i.e., the angle between the cone's axis and the
$\eta$-axis is zero); then, the cone is enlarging and inclining
so that in the future (see Eqs.(\ref{coneout})) its angle becomes equal
to $\alpha_1=\pi-\arctan(1+\aout)-\arctan(1-\aout)$ and the
inclination's angle becomes equal to
$\alpha_2=\frac12[\arctan{(1+\aout)}-\arctan{(1-\aout)}]$.  There are
two qualitatively different cases: (i) $\aout<1$ and (ii) $\aout\ge
1$. In the first case, $\aout<1$, this rotation of the cone on the
$\eta$-$\xi$ plane does not lead to the appearance of causal
pathologies, i.e. all future-directed world lines remain unclosed. In
the second case, $\aout\ge 1$, the situation is cardinally changed
(this case is illustrated by the spacetime diagram in the figure 1).

\begin{figure}[t]
\epsffile{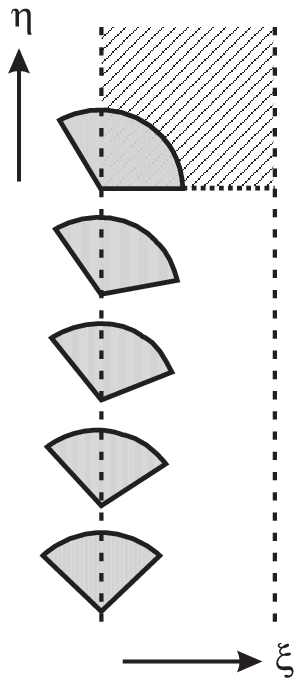}   
\begin{center} Fig.1. \end{center}
\end{figure}

\noindent
Namely, now there is such a moment of time (the value of
$\eta=\eta_{*}$) when one of the cone's side takes a `horizontal'
position, i.e. one of the null curve's equations becomes $\eta=const$.
But these lines are closed (see the Eq.(\ref{ident})), and hence, at
the moment $\eta_*$ the closed null curves appear in our model. We
shall speak that a time machine is being formed at this moment of time.
If $\eta_*=\infty$ ($\aout=1$) the time machine is formed in the 
infinitely far
future. Otherwise, it is formed at the time $\eta_*<\infty$. In this
case at later times $\eta>\eta_*$ the closed line $\eta=const$ lies
{\it inside} of the light cone, i.e.  the region $\eta>\eta_*$ contains
closed timelike curves.

Thus we may conclude we have constructed the model of spacetime in
which, depending on the parameter $\aout$, the chronology horizon is
being formed at some moment of time.

Now let us investigate a behavior of quantized fields in this model. 

\section{A particle creation}

\subsection{Scalar field: solutions of the wave equation}

Consider a conformal massless scalar field $\phi$ for which the
Lagrangian is
\beq\label{lagr}
{\cal L}= \frac12\nabla_{\mu}\phi\nabla^{\mu}\phi.
\eeq
The scalar field $\phi$ obeys the wave equation
\beq\label{fe}
\Box\phi=0.
\eeq
In addition, it follows from the identification rule (\ref{ident}) and
from the quadric form of the lagrangian (\ref{lagr}) that
the scalar field has to obey the periodic condition (the ordinary
field) 
\beq\label{percond}
\phi(\eta,\xi+L)=\phi(\eta,\xi)
\eeq
or the antiperiodic one (the twisted field)
\beq\label{antipercond}
\phi(\eta,\xi+L)=-\phi(\eta,\xi).
\eeq
In the metric (\ref{M}) the wave equation (\ref{fe}) reads
\beq\label{FE}
\Big[ (1-a^2)\partial_\eta^2+2a\partial_\eta\partial_\xi-
\partial_\xi^2-2aa'\partial_\eta+a'\partial_\xi \Big] \phi(\eta,\xi) =0,
\eeq
where \mbox{$\partial_\eta=\frac{\partial}{\partial\eta}$}, 
\mbox{$\partial_\xi=\frac{\partial}{\partial\xi}$} and a prime denotes the
derivative on $\eta$, $a'=da/d\eta$.

First of all, let us solve this equation in the asymptotical regions.
In the in-region, where $a(\eta)\to 0$ and $a'(\eta)\to 0$, the
Eq.(\ref{FE}) reduces to  
\beq
[ \partial_\eta^2 - \partial_\xi^2 ] \phi(\eta,\xi) =0,
\eeq
The complete set of solutions of this equation is
\beq\label{inmodes}
\phi_n^{(\pm,\rm in)}=D_n^{\rmin} 
e^{ik_n\xi}e^{\mp i\omega\eta}, 
\eeq
where 
$$k_n=\frac{2\pi n}{L},\ n=\pm1,\pm2,\dots\quad\rm (ordinary~field)$$
or
$$k_n=\frac{2\pi{\textstyle (n+\frac12)}}{L},\ n=0,\pm1,\pm2,\dots\quad\rm
(twisted~field),$$
$\omega=|k_n|$ and $\phi_n^{(+,\rm in)}$ and $\phi_n^{(-,\rm
in)}$ are positive and negative frequency solutions (the ``in-modes'')
in the in-region, respectively. The in-modes form the basis
in Hilbert space $\cal H$ with the scalar product
\beq\label{sp}
(\phi_n, \phi_{n'})=-i\int(\phi_{n} \,
{\mathop{\partial_\mu}\limits^\leftrightarrow } \,\phi^*_{n'}) 
d\Sigma^\mu,
\eeq
where $d\Sigma^\mu = d\Sigma\, n^\mu$, with $d\Sigma$ being the volume
element in a given spacelike hypersurface, and $n^\mu$ being the
timelike unit vector normal to this hypersurface. 
Choosing the hypersurface $\eta=const$ we may write down the scalar
product (\ref{sp}) in the in-region as follows:
\beq\label{sp:in}
(\phi_n, \phi_{n'})=-i\int_0^L d\xi
\left(\phi_n\frac{\partial\phi^*_{n'}}{\partial\eta}-
\frac{\partial\phi_{n}}{\partial\eta}\phi^*_{n'}\right)_{\eta=const}
\eeq
The in-modes are orthonormal provided
\beq\label{Dnin}
D_n^\rmin=\frac{1}{\sqrt{4\pi |n+\alpha|}},
\eeq
where $\alpha=0$ for an ordinary scalar field and $\alpha=\frac12$ for a
twisted one.

Analogously we may find a solution of the wave equation
(\ref{FE}) in the out-region. There $a(\eta)\to a_{\rm o}$ and
$a'(\eta)\to 0$, and the Eq.(\ref{FE}) reduces to 
\beq
\Big[ (1-\aout^2)\partial_\eta^2+2\aout\partial_\eta\partial_\xi-
\partial_\xi^2\Big] \phi(\eta,\xi) =0,
\eeq
The complete set of solutions of this equation is  
\beq\label{outmodes}
\phi_n^{(\pm,\rm out)}= D_n^{\rm (out)} e^{ik_n\xi}
e^{-ik_na_{\rm o}\beta^{-1}\eta}e^{\mp i\omega\beta^{-1}\eta},
\eeq
where we denote 
\beq\label{beta}
\beta=1-\aout^2,
\eeq
and $\phi_n^{(+,\rm out)}$ and $\phi_n^{(-,\rm out)}$ are positive and
negative frequency solutions (the ``out-modes'') in the out-region,
respectively. The out-modes (\ref{outmodes}) form the basis
in Hilbert space $\cal H$ with the scalar product (\ref{sp}) which in
the out-region reads
$$
(\phi_n, \phi_{n'})=-i\int_0^L d\xi
\left\{\left(\phi_n\frac{\partial\phi^*_{n'}}{\partial\eta}-
\frac{\partial\phi_{n}}{\partial\eta}\phi^*_{n'}\right)\right.
$$
\beq\label{sp:out}
+\left.\frac{\aout}{1-\aout^2}
\left(\phi_n\frac{\partial\phi^*_{n'}}{\partial\xi}-
\frac{\partial\phi_{n}}{\partial\xi}\phi^*_{n'}\right)\right\}
_{\eta=const}
\eeq
The out-modes are orthonormal provided
\beq\label{Dnout}
D_n^{\rm (out)}=\sqrt{\frac{\beta}{4\pi |n+\alpha|}},
\eeq

Now let us solve the Eq.(\ref{FE}) in a general case.  Noting that the
metric (\ref{M}) is invariant under translations in the $\xi$-direction
and taking into account the periodic or antiperiodic conditions
(\ref{percond}), (\ref{antipercond}) we may find solutions in the
following form:
\beq\label{phi}
\phi_n(\eta,\xi)=u_n(\eta)e^{i k_n\xi}.
\eeq
Substituting the expression (\ref{phi}) into the
wave equation (\ref{FE}) we obtain the equation for $u_n(\eta)$:
\beq\label{u}
(1-a^2)u''+2a(ik_n-a')u'-ik_n(ik_n-a')u=0.
\eeq
Introduce a new function $v(\eta)$ by the relation
\beq\label{defv}
u(\eta)=v(\eta)
\exp{\left(-\int^\eta\frac{a(ik_n-a')}{1-a^2}d\tilde\eta\right)}.
\eeq
Note that in the in-region the relation (\ref{defv}) has the
asymptotical form
\beq\label{uin}
u(\eta)=v(\eta),
\eeq
whereas in the out-region it reads
\beq\label{uout}
u(\eta)=v(\eta)e^{-ik_n\aout\beta^{-1}\eta}
\eeq
After substituting Eq.(\ref{defv}) into Eq.(\ref{u}) we obtain the
second-order differential equation for $v(\eta)$ in a so-called normal
form $v''+\Omega^2(\eta)v=0$:
\beq\label{v}
v''+\left(\frac{k_n^2+a'^2+a''a(1-a^2)}{(1-a^2)^2}\right)v=0.
\eeq
Further we shall solve this equation {\it only}\ for modes for which the
conditions 
\beq\label{condi}
k_n^2>\!\!>a'^2,\quad k_n^2>\!\!>a''a(1-a^2)
\eeq
are fulfilled. That is we shall only consider the modes whose wave
length is much less than a typical scale of variation of the function
$a(\eta)$. Taking into account the conditions (\ref{condi}) we could
now neglect in (\ref{v}) the terms $a'^2$ and $a''a(1-a^2)$ and 
rewrite\footnote[1]{Let me emphasize that the approximation which we use
is no WKB approximation. Remind that the last means, roughly speaking(!),
neglecting terms with $\Omega'$, $\Omega''$, $\dots$ in the solution
of the equation $v''+\Omega^2(\eta)v=0$.} 
\beq\label{vapprox}
v''+\frac{k_n^2 v}{(1-a^2)^2}=0.
\eeq
Now let us restrict our consideration by the {\it special form} of the
function $a(\eta)$: 
\beq\label{a}
a^2(\eta)=\frac12\aout^2(1+\tanh\gamma\eta),
\eeq
where $\gamma$ is a parameter. It is easy to see that the function
$a(\eta)$ defined by Eq.(\ref{a}) possesses the necessary asymptotical
behavior (\ref{a_asymp}). The quantity $T=(2\gamma)^{-1}$ gives
the typical time of variation of the function $a(\eta)$ from one 
asymptotical value to another one. The conditions (\ref{condi}) reduce
now to
\beq
k_n^2>\!\!> \frac{4}{27}\frac{\aout^2}{T^2}.
\eeq

Substituting the expression (\ref{a}) into Eq.(\ref{vapprox}) we could
rewrite the equation (\ref{vapprox}) as
\beq\label{va}
v''+\frac{4k_n^2 v}{(2-\aout^2(1+\tanh\gamma\eta))^2}=0.
\eeq
A solution of this equation could be found in terms of hypergeometric 
series. In paticular, to find a solution which will be positive
frequency in the in-region we have to choose the solution of the
equation (\ref{va}) as follows: 
\beq\label{solin}
v_n^\rmin (\eta)=D_n^\rmin(-z)^{-i\mu}(1-\beta z)^\sigma
F(q,r;s;\beta z),
\eeq
where $D_n^\rmin$ is defined by (\ref{Dnin}) and
\beq\label{abc}
\begin{tabular}{l}
$\displaystyle
z=-e^{2\gamma\eta},\quad \mu=\frac{\omega}{2\gamma},\quad 
\beta=1-\aout^2 $\\[3pt]
$\displaystyle s=1-2i\mu,\quad 
\sigma=\frac12+\sqrt{\frac14-\mu^2\frac{(1-\beta)^2}{\beta^2}}$\\[3pt]
$\displaystyle q=\sigma+i\mu\frac{1-\beta}{\beta},\quad
r=\sigma-i\mu\frac{1+\beta}{\beta},$
\end{tabular}
\eeq
and $F(q,r;s;z)$ is a Gaussian hypergeometric function. Taking into
account the relations (\ref{defv}) and (\ref{phi}) we can now write
down the solution of the wave equation (\ref{FE}) as 
\beqa\label{phiin} \nonumber
f_n(\eta,\xi)&=&D_n^\rmin e^{ik_n\xi}e^{-i\omega\eta}
\exp{\left(-\int^\eta\frac{a(ik_n-a')}{1-a^2}d\tilde\eta\right)}
\\ 
&&\times(1-\beta z)^\sigma F(q,r;s;\beta z)
\eeqa
Note that $F(q,r;s;0)=1$ for any $q$, $r$ and $s$. Now it is not difficult
to see that this solution in the in-region, where $\eta\to-\infty$ or
$z\to-0$, has the following asymptotical form: 
$f_n(\eta,\xi)\approx D_n^\rmin e^{ik_n\xi}e^{-i\omega\eta}$,
which coincides with the positive frequency in-modes 
$\phi_n^{(+,\rm in)}$. 

\subsection{Bogolubov coefficients}

Now let us remind some mathematical aspects for describing of the
physical phenomenon of partical creation by a time-depend gravitational
field. 

So, we have obtained the set of solutions $\{f_n\}$ which are positive
frequency (the ``in-modes'') in the past. Let $\{F_n\}$ be positive
frequency solutions (the ``out-modes'') in the future. (We do not need
to know an explicit form of these solutions. It will be enough to know
their asymptotic properties in the out-region, i.e.
$F_n\approx\phi_n^{(+,\rm out)}$.) We may choose these two
sets of solutions to be orthonormal, so that
\beqa\label{ortho}
&(f_n,f_{n'})= (F_{n},F_{n'})=\delta _{nn'}& \nonumber \\
&(f_n^*,f_{n'}^*)= (F_{n}^*,F_{n'}^*)= -\delta _{nn'}& \nonumber \\
&(f_n,f_{n'}^*)= (F_{n},F_{n'}^*)= 0.&  
\eeqa
The in-modes may be expanded in terms of the out-modes:
\beq\label{expand}
f_n=\sum\limits_m {(\alpha _{nm}}F_m + \beta _{nm}F_m^*).
\eeq
Inserting this expansion into the orthogonality relations, 
Eq. (\ref{ortho}), leads to the conditions
\beq\label{condab}
\sum\limits_m {(\alpha _{nm}\alpha _{n'm}^*-\beta _{nm}\beta_{n'm}^*)
= \delta _{nn'}}, 
\eeq
and
\beq
 \sum\limits_m (\alpha _{nm}\alpha _{n'm}-\beta _{nm}\beta _{n'm})=0.
\eeq
The field operator, $\phi$, may be expanded in terms of either the
$\{ f_n \}$ or the $\{ F_n \}$:
\beq
  \phi = \sum\limits_n (a_n f_n + a_n^\dagger f_n^*)
  =\sum\limits_n (b_n F_n + b_n^\dagger F_n^*).
\eeq
The $a_n$ and $a_n^\dagger$ are annihilation and creation operators,
respectively, in the in-region, whereas the $b_n$ and $b_n^\dagger$ are
the corresponding operators for the out-region. The in-vacuum state is
defined by $a_n|0\rangle_{in}=0, \; \forall n,$ and describes the situation
when no particles are present initially. The out-vacuum state is
defined by $b_n|0\rangle_{out}=0, \; \forall n,$ and describes the
situation when no particles are present at late times. Noting that $a_n
= (\phi,f_n)$ and $b_n = (\phi,F_n)$, we may expand the two sets of
creation and annihilation operator in terms of one another as 
\beq\label{bog1}
a_n=\sum\limits_m (\alpha _{nm}^*b_m-\beta _{nm}^* b_m^\dagger),                                   
\eeq
or 
\beq\label{bog2}
b_m=\sum\limits_n (\alpha _{nm} a_n + \beta _{nm}^* a _n^\dagger).                                      
\eeq
This is a Bogolubov transformation, and the $\alpha_{nm}$ and
$\beta_{nm}$ are called the Bogolubov coefficients.

Let us assume that no particle were present before the gravitational
field is turned on. In the Heisenberg approach $|0\rangle_{in}$ is the
state of the system for all time. However, the physical number operator
which counts particles in the out-region is $N_m = b_m^\dagger b_m$.
Thus the number of particles created into mode $m$ is
\beq
\langle N_m \rangle = {}_{in}\langle 0|b_m^\dagger b_m |0\rangle_{in}
       = \sum\limits_n |{\beta _{nm}}|^2.
\eeq

To find the Bogolubov coefficients in our case we have to determine an 
asymptotical form of the in-modes in the out-region where
$\eta\to\infty$ or $z\to-\infty$. With this aim we consider
the analitical continuation of the hypergeometric function $F(q,r;s;z)$
into the region of large values of $|z|$ \cite{AS}
\beqa\label{ancont} 
\nonumber&&F(q,r;s;z)= \\[3pt]
\nonumber&&\displaystyle\phantom{+}
\frac{\Gamma(s)\Gamma(r-q)}{\Gamma(r)\Gamma(s-q)} 
(-z)^{-q}F(q,1-s+q;1-r+q;\frac1z)\\[3pt]
&&\displaystyle+\frac{\Gamma(s)\Gamma(q-r)}{\Gamma(q)\Gamma(s-r)}
(-z)^{-r}F(r,1-s+r;1-q+r;\frac1z) .
\eeqa
After substituting this expression into the Eq.(\ref{phiin}) it is not
difficult to see that the asymptotic of the in-modes in the out-region
is 
\beqa
\nonumber&&f_n(\eta,\xi)\approx D_n^\rmout
e^{ik_n\xi}e^{-ik_n\aout\beta^{-1}\eta} \\[3pt]
\nonumber&&\displaystyle\times\left( (-\beta)^{-i\mu\beta^{-1}(1-\beta)}
\frac{\Gamma(s)\Gamma(r-q)}{\beta^{1/2}\Gamma(r)\Gamma(s-q)}
e^{-i\omega\beta^{-1}\eta} \right. \\[3pt]
&&\displaystyle+\left. (-\beta)^{i\mu\beta^{-1}(1+\beta)}
\frac{\Gamma(s)\Gamma(q-r)}{\beta^{1/2}\Gamma(q)\Gamma(s-r)}
e^{i\omega\beta^{-1}\eta} \right) .
\eeqa
Comparing the last expression with the asymptotical form of the
out-modes, Eq.(\ref{outmodes}), we may write
\beqa \nonumber
&\phi_n^\rmin (\eta,\xi)&\ = A_n \phi_n^{(+,\rm out)}+
B_n \phi_{n}^{*(+,\rm out)}\\
&&\ = A_nF_n+B_nF_{-n}^{*}
\eeqa
where 
\beq
A_n=(-\beta)^{-i\mu\beta^{-1}(1-\beta)}
\frac{\Gamma(s)\Gamma(r-q)}{\beta^{1/2}\Gamma(r)\Gamma(s-q)},
\eeq
\beq\label{Bn}
B_n=(-\beta)^{i\mu\beta^{-1}(1+\beta)}
\frac{\Gamma(s)\Gamma(q-r)}{\beta^{1/2}\Gamma(q)\Gamma(s-r)},
\eeq
As follows from Eq.(\ref{expand}), the coefficients $A_n$ and $B_n$ are
the related to the Bogolubov coefficients by
$\alpha_{nm}=A_n\delta_{nm}$, $\beta_{nm}=B_n\delta_{n,-m}$. The number
of particles created into mode $n$ is now determined as
\beq
\langle N_n \rangle =  |B_n|^2.
\eeq
Using the Eq.(\ref{Bn}) gives
\beq
\langle N_n \rangle =  \beta^{-1}
\left|\frac{\Gamma(s)\Gamma(q-r)}{\Gamma(q)\Gamma(s-r)}\right|^2
\eeq
To carry out calculations in the above expression we shall use the
following formulae \cite{AS}:
\beqa\nonumber
&&|\Gamma(iy)|^2=\frac{\pi}{y\sinh \pi y},\quad 
|\Gamma({\textstyle\frac12+iy})|^2=\frac{\pi}{\cosh \pi y},\\
&&|\Gamma(1+iy)|^2=\frac{\pi y}{\sinh \pi y}.
\eeqa
By using this formulae and the relations (\ref{abc}) we may finally
obtain the spectrum of created particles:
\beq\label{spectrum}
\langle N_n \rangle=
\frac{\cosh\frac{\pi\omega(1-\beta)}{\gamma\beta}+
\cosh\pi\sqrt{\big(\frac{\omega(1-\beta)}{\gamma\beta}\big)^2-1}}
{2\sinh\frac{\pi\omega}{\gamma}\sinh\frac{\pi\omega}{\gamma\beta}}.
\eeq

\subsection{Particle creation near the chronology horizon}

Now let us analyse how much particles of the massless scalar field is
created at the moment of the chronology horizon forming. As was
mentioned above, the value of the parameter $\aout$ determines either
the time machine is formed or not. If $\aout<1$ then the chronology
horizon is absent in the spacetime, but if $\aout=1$ then it appears in
the infinitely far future. The expression (\ref{spectrum}) gives us the
spectrum of particles created in the out-region. The spectrum depends
on the parameter $\aout$ ($\beta=1-\aout^2$), and in the limit
$\aout\to 1$ the expresion (\ref{spectrum}) will determine the number
of particles created near the chronology horizon. Going to the limit
$\aout\to 1$ in Eq.(\ref{spectrum}) gives
\beq\label{limspectr}
\langle N_n \rangle=\frac{1}{\sinh\frac{\pi\omega}{\gamma}}.
\eeq
Thus we see that the number of particles created into mode $n$ near the
chronology horizon is finite. We may also conclude that the full number
of particles $N=\sum_n\langle N_n \rangle$ will be finite
because the spectrum (\ref{limspectr}) is exponentially decreasing.

\section{Conclusion}
Let us summarize. In this work we have constructed the model of
spacetime in which, depending on the model's parameter $\aout$, the
chronology horizon could be formed; there are no causal pathologies if
$\aout<1$, and the chronology horizon appears at the moment of time
$\eta_*<\infty$ if $\aout>1$. In the case $\aout=1$ closed lightlike
curves are formed in the infinitely far future. The model represents a
two-dimensional curved spacetime with the topology $R^1 \times S^1$
which is asymptotically flat in the past and in the future, but which
is non-flat in the intermediate region. As a consequence, in this
spacetime the creation of particles of quantized fields by the
gravitational field is possible. We have studied the particle creation
of a massless scalar field in the case $\aout<1$, i.e. in the case when
the spacetime is globally hyperbolic and has no causal pathologies.  As
a result, the spectrum of created particles has been obtained (see
Eq.(\ref{spectrum})).  In the limit $\aout\to1$ this spectrum gives the
number of particles created into mode $n$ near the chronology horizon
(see Eq.(\ref{limspectr})). The main result we have obtained is that
the number of scalar particles created into each mode as well as the
full number of particles remain finite at the moment of forming of the
chronology horizon. This result might mean that the phenomenon of
particle creation could not prevent the formation of time machine.
However, to do the final conclusion one has to take into account a
backreaction of created particles on a spacetime metric.

\section*{Acknowledgment}
I would like to thank my colleague Sosov E. N. for helpful discussions. 
This work was supported in part by the Russian Foundation for Basic
Research under grant No. 96-02-17066.


\begin{references}
\bibitem{Refs}
The reader could find a review of various aspects of the time
machine problem as well as a lot of references in an exellent monograph
\cite{VisserBook}
\end{references}
\end{document}